\documentclass[aps,prl,twocolumn,amsmath,floatfix]{revtex4-2}

\usepackage{graphicx} 
\usepackage{braket}
\usepackage[colorlinks,allcolors=blue]{hyperref}

\usepackage{color}

\begin{document}
\title{Localization of light in three dimensions:\\
a mobility edge in the imaginary axis in non-Hermitian Hamiltonians}

\author{Giuseppe Luca Celardo}
\affiliation{Department of Physics and Astronomy, CSDC and INFN, Florence Section, University of Florence, Italy. European Laboratory for Non-Linear Spectroscopy, Università di Firenze, Via Nello Carrara 1, 50019 Sesto Fiorentino, Italy.}
\author{Mattia Angeli}
 \affiliation{John A. Paulson School of Engineering and Applied Sciences, Harvard University, Cambridge, Massachusetts 02138, USA}
\author{Francesco Mattiotti}
\affiliation{CESQ and ISIS (UMR 7006), University of Strasbourg and CNRS, 67000 Strasbourg, France}
\author{Robin Kaiser}
\affiliation{Universit\'e C\^ote d'Azur, CNRS, Institut de Physique de Nice, 06200 Nice, France.}
\date{\today}

\begin{abstract}
Searching for Anderson localization of light in three dimensions has challenged experimental and theoretical research for the last decades. Here the problem is analyzed through large scale numerical simulations, using a radiative Hamiltonian \emph{i.e.}~a non-Hermitian long-range hopping Hamiltonian, well suited to model light-matter interaction in  cold atomic clouds. 
 Light interaction in atomic clouds is considered in presence of positional and diagonal disorder.  Due to the interplay of disorder and cooperative effects (sub- and super-radiance)  a novel type of localization transition is shown to emerge, differing in several aspects from standard localization transitions which occur along the real energy axis. The localization transition discussed here is characterized by a mobility edge along the imaginary energy axis of the eigenvalues which is mostly independent from the real energy value of the eigenmodes. Differently from usual mobility edges it  separates extended states from hybrid localized states and it manifest itself in the large moments of the participation ratio of the eigenstates.  Our prediction of a mobility edge in the imaginary axis, \emph{i.e.}~depending on the eigenmode lifetime, paves the way to achieve control both in the time and space domain of open quantum systems.
\end{abstract}  

        
\maketitle

\paragraph{Introduction.}

The interplay of opening and disorder in systems described by non-Hermitian Hamiltonians has been at the center of interest in many research fields, showing that non-Hermiticity can strongly affect the response of a system to disorder, inducing many counter-intuitive effects~\cite{EichelkrautNatCom2013,LuoPRL2021,KawabataPRL2021,huangPRB2020,WeidemannNatPhot2021,Celardo13a,Celardo13b,robin2019,Giulio14,Giulio15,SR2c}. On the other side, Anderson localization~\cite{Anderson58} has been a beacon to understand closed disordered systems and has been at the focus of an ever increasing research community, ranging from condensed matter to acoustics, optics, and ultra-cold matter waves as well as quantum memories based on cold atoms~\cite{Ramakrishnan,Akkermans,Lagendijk09,Aspect09,Page08,Garreau08,DeMarco11,Aspect12,Modugno15}. In the standard Anderson localization problem, an excitation can tunnel to nearest-neighbor sites placed in a regular lattice with disordered on-site energies (diagonal disorder). Depending on the value of the disorder strength, a mobility edge can be present at a specific energy: below this energy the eigenstates are localized, while above they are delocalized.

\begin{figure}[!htb]
\centering
\includegraphics[width=2.7in]{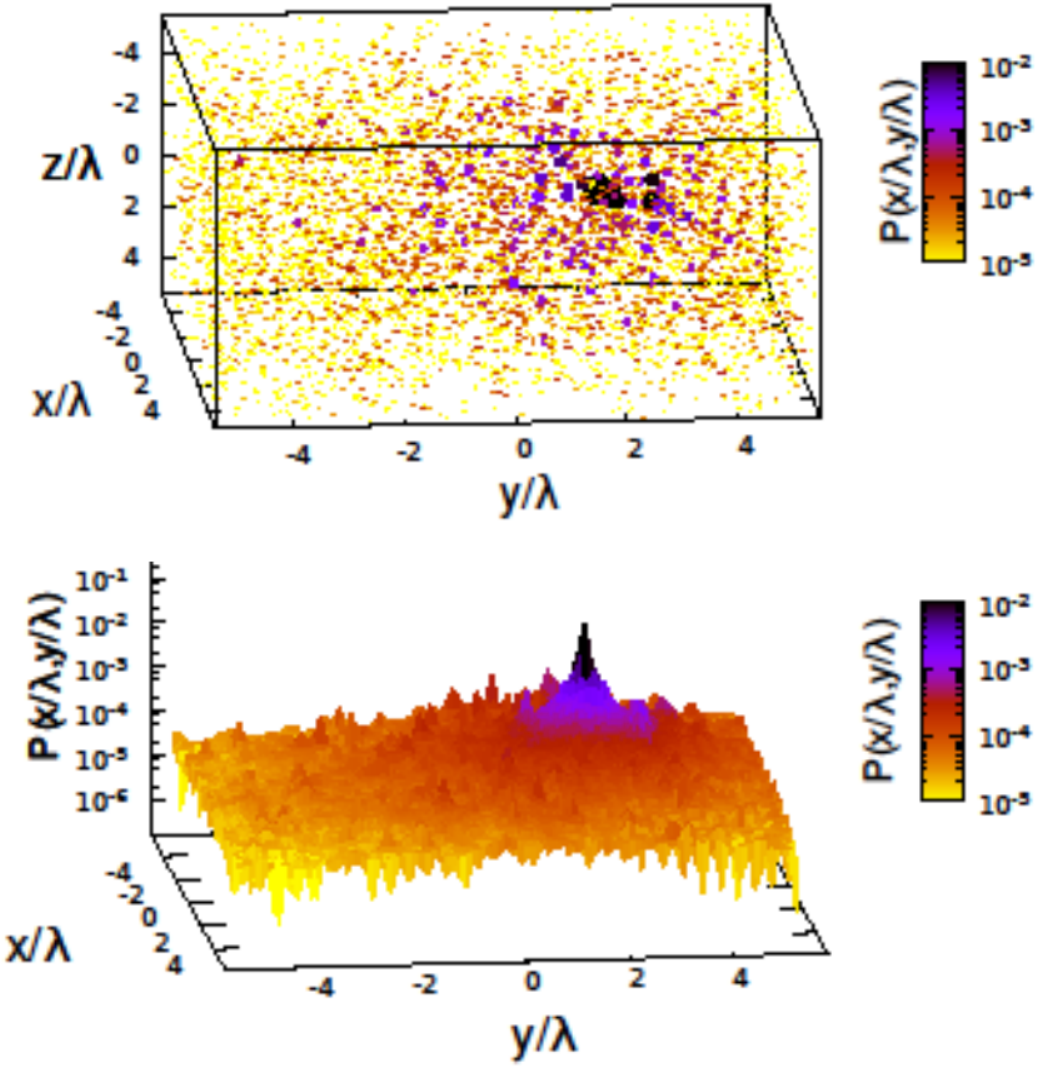}  \\
\caption{(Color online) {\it Localized Subradiant Eigenstates}.
A typical localized subradiant state for the scalar model and for $\Gamma=0.094 \Gamma_0, E=0.1 \Gamma_0$ and a participation ratio $PR_2\approx 7$ for the case  $W/(\Gamma_0 b_0)=0.4$ and $N=6400,\rho \lambda^3=5$ so that $b_0 \approx 17.3$.  Upper panel: Three-dimensional representation of a localized eigenstate. The radius representing each atom is proportional to its excitation probability  $|\Psi_j(r)|^2$, also coded in color~\cite{sp}.
  Lower panel: A localized eigenstate projected on the $x-y$ plane.
}
\label{fig:Fig1}
\end{figure}

Extending the concepts developed for Anderson localization to open quantum systems still remains a challenge. Light has been an obvious candidate to study Anderson localization of non-interacting waves, which has triggered continuous efforts since the mid-80s~\cite{John84,John86,Anderson85,Lagendijk97,Maret06,Maret13,Scheffold99,Lagendijk12,Scheffold13,Maret16,Skipetrov16}. So far, Anderson localization of light in three dimensions however has resisted experimental observation. It has now been shown that pioneering experiments on Anderson localization of light~\cite{Lagendijk97,Maret06,Maret13} do not provide a signature for the Anderson transition in three dimensions~\cite{Scheffold99,Lagendijk12,Scheffold13,Maret16, Skipetrov16} and the mere existence of an Anderson phase transition for light had even been questioned~\cite{Skipetrov14,Bellando14}.
Localization of light indeed presents many features which strongly differ from the standard Anderson localization of closed systems: $i)$ in typical samples, scatterers have random positions in a three-dimensional volume, leading to positional disorder, $ii)$ light induces complex long-range hopping  between the sites, which in the case of two-level systems as scattering medium can lead to cooperative effects such as Dicke sub- and superradiance~\cite{Kaiser08,Kaiser09,Kaiser12,Akkermans13,Guerin16c}, $iii)$ the excitation can escape from the system by photon emission, thus placing the problem of localization of light within the framework of open quantum systems. 
Both the long-range nature of the hopping and the opening can strongly affect the interplay of disorder and transport. Thus, the possibility to have a  transition to localization in such systems is highly non-trivial.
Specifically, cooperativity  can affect the response of the system to disorder in a drastic way: while
superradiant states show robustness to disorder~\cite{Giulio14,Giulio15}, in the subradiant subspace 
long-range interaction is effectively shielded~\cite{Celardo16,lea} and signatures of localization can emerge~\cite{Celardo16,Celardo13a,Celardo13b}.  
In this letter, we shed new light on the problem of light localization in resonant scattering media by combining the positional disorder studied so far, with the initial ingredient of diagonal (on-site) disorder 
as
considered by Anderson~\cite{Anderson58}. With the aid of large scale simulations of up to 50000 atoms relying on a well-known radiative non-Hermitian Hamiltonian~\cite{mattiotti2020nano} able to take the vectorial nature of light into account,  we show that the playground of the problem of light localization lies in the complex eigenvalues of the radiative Hamiltonian. Specifically, by analyzing the generalized participation ratio (GPR) of the eigenstates, a widely used figure of merit to study localization transitions, we show that a drastic transition in the behaviour of the GPR occurs at a specific imaginary energy value (decay widths) of the eigenstates. Such transition shares many analogies with what happens at real energy mobility edges and thus reveals the presence of a mobility edge along the imaginary axis in such systems.  

\paragraph{The Model.} 
We model light scattering in a 3D cold atomic cloud by considering $N$ atoms randomly distributed inside a
cube of volume $V={L}^3$, with a spatial density $ \rho=N/ {L}^3$. When considering the interaction of atoms with the electromagnetic field, 
the full vectorial character of light should be taken into account. We  will focus on  atoms driven on a $s\rightarrow p$ dipole radiation transition which can be characterized by three degenerate levels in the excited state (labelled as $\alpha=x,y,z$), each with a transition dipole moment (TDM) equal in coupling strength and perpendicular to the others~\cite{Bellando14}.
Thus we model each atom as a four-level system, with a ground state $\ket{g}$ and three degenerate excited states $\ket{x}$, $\ket{y}$ and $\ket{z}$.
Corresponding TDM matrix elements are $\langle\alpha|\widehat{\vec{\mu}}|g\rangle=\mu{\hat{e}}_\alpha$, with $\alpha=x,y,z$ and the Cartesian unit vectors defined as ${\hat{e}}_\alpha$.
The radiative hamiltonian $\mathcal{H}_{\rm vec}$ describing dipole-dipole coupling in the single excitation approximation (see also~\cite{mattiotti2020nano}) is
\begin{multline}
\label{hamvec}
  \mathcal{H}_{\rm vec} = \sum_{n=1}^N\sum_{\alpha\in\{x,y,z\}} \left( E_{n,\alpha} - i\frac{\Gamma_0}{2} \right) \ket{n,\alpha} \bra{n,\alpha} \\
  - \frac{\Gamma_0}{2} \sum_{\substack{m,n=1 \\ \left( m \neq n \right)}}^N \sum_{\alpha,\beta \in \{x,y,z\}} V_{m,n,\alpha,\beta}\ket{m,\alpha}\bra{n,\beta}~,
\end{multline}
where $E_{n,\alpha}$ are the atomic transition energies and $\Gamma_0$ is the radiative decay rate of a single atom. In Eq.~\eqref{hamvec}, $\ket{n,\alpha}$ represents a quantum state where the $n$th atom is excited in its $\alpha$th state, while all the other atoms are in their ground state. Interaction terms are non-Hermitian, namely
\begin{widetext}
\begin{align}
\label{hamc}
V_{m,n,\alpha,\beta} &= \frac{3}{2} e^{i k_0 r_{m,n}} \left[ \left( \frac{1}{k_0 r_{m,n}} + \frac{i}{k_0^2 r_{m,n}^2} - \frac{1}{k_0^3 r_{m,n}^3} \right) \hat{e}_\alpha \cdot \hat{e}_\beta - \left( \frac{1}{k_0 r_{m,n}} + \frac{3i}{k_0^2 r_{m,n}^2} - \frac{3}{k_0^3 r_{m,n}^3} \right) (\hat{e}_\alpha \cdot \hat{r}_{m,n}) (\hat{e}_\beta \cdot \hat{r}_{m,n}) \right].
\end{align}
\end{widetext}
In Eq.~\eqref{hamc},
$k_0=2\pi/\lambda=E_0/(\hbar c)$ is the transition wavenumber (where $\lambda$ is the wavelength of the atomic transition and $E_0$ being the average single atomic transition energy $E_0=\braket{E_{n,\alpha}}$, where the average is taken over disorder realization. $r_{m,n}$ is the distance between the $m$th and $n$th atom and ${\hat{r}}_{m,n}$ is the unit vector joining them.
Together with the vectorial model we also consider the scalar model~\cite{Cipris21}.
Even though the latter approximation neglects polarisation effects, it is appropriate in the dilute limit, where inter-atomic distances are larger than the optical wavelength $\lambda$, making near-field terms decaying as $1/r^3$ negligible~\cite{sp}. 
The effective Hamiltonian which governs the interaction of the atoms with the electromagnetic field in the scalar approximation is characterized by complex long-range hopping terms $V_{m,n}$ decreasing as $1/r_{m,n}$ with the distance,
\begin{equation}
\mathcal{H}=  \sum_{n=1}^N \left(E_n - i\frac{\Gamma_0}{2}\right) \ket{n}\bra{n}-\frac{\Gamma_0}{2}
\sum_{m\neq n}^N V_{m,n} \ket{m}\bra{n},
\label{Hscalar}
\end{equation}
where the state $\ket{n}$ stands for the $n-$atom in the excited state and  
all the other atoms being in the ground state, while $V_{m,n}=\frac{\exp(ik_0  r_{m,n})}{k_0 r_{m,n}} $ is the interaction
between the atoms at distance $r_{m,n}$. The model in Eq.~(\ref{Hscalar}), known as the scalar model, has been introduced first by Foldy~\cite{foldyscalar} and it has been used in several papers to describe cold atomic clouds in the dilute limit~\cite{Guerin16a}. 
Note that the vectorial model Hamiltonian has dimension $3N \times 3N$ contrary to the scalar case which has dimension $N \times N$. Thus the scalar model allows us to investigate much larger system sizes with better statistics.
For both models we can define the resonant mean free path $l=1/\rho\sigma_0$ (in the independent scattering approximation), where  $\sigma_0=4\pi/k_0^2$ is the resonant scattering cross section in a simplified scalar model.
Finally, we define the resonant optical thickness, $b_{0}$, as the ratio
between the system size $L$ and the mean free path $l$. For the scalar model we have
\begin{equation}
b_{0}= \frac{L}{l}=\frac{4 \pi \rho^{2/3} N^{1/3}}{k_0^2},
\label{b0}
\end{equation}
 while  for the vectorial model the resonant scattering cross section is $\sigma_0=6\pi /k_0^2$ and the optical thickness thus has to be corrected with respect to the scalar case :  $b_0^{\rm (vec)}=(3/2) b_0^{\rm (scal)}$.

\begin{figure}[h!]
\centering
\includegraphics[width=2.6in]{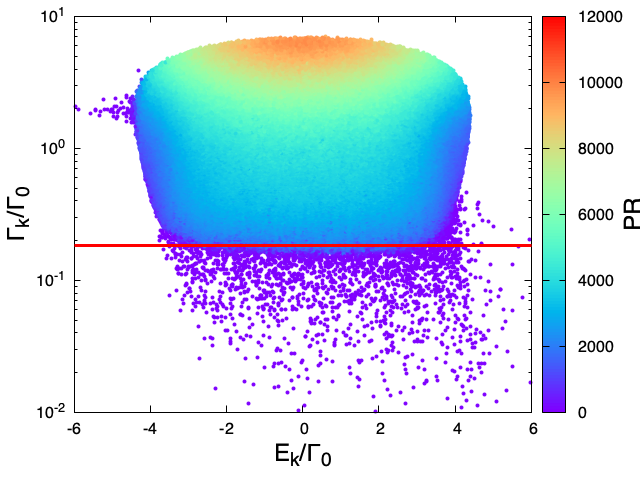} 
\caption{(Color online)  {\it Participation ratio in the complex plane}: The participation ratio $PR=PR_{q=2}$ of the eigenstates for the scalar model is shown in the complex plane $(E_k,\Gamma)$ of the complex eigenvalues  for $N=28^3$, $\rho\lambda^3=5$, $b_0 \approx 26.06$ and  $W/(b_0 \Gamma_0)=0.3$. Note that the horizontal energy axis is shifted by $E_0$. 
The critical width for the transition to localization [Eq.~(\ref{Gcr})] is indicated by the red  horizontal line. Note that $E_k$ is the difference between the real part of the eigenvalues and $E_0$.    }
\label{fig:Fig2a}
\end{figure}

\begin{figure*}
\centering
\includegraphics[width=\textwidth]{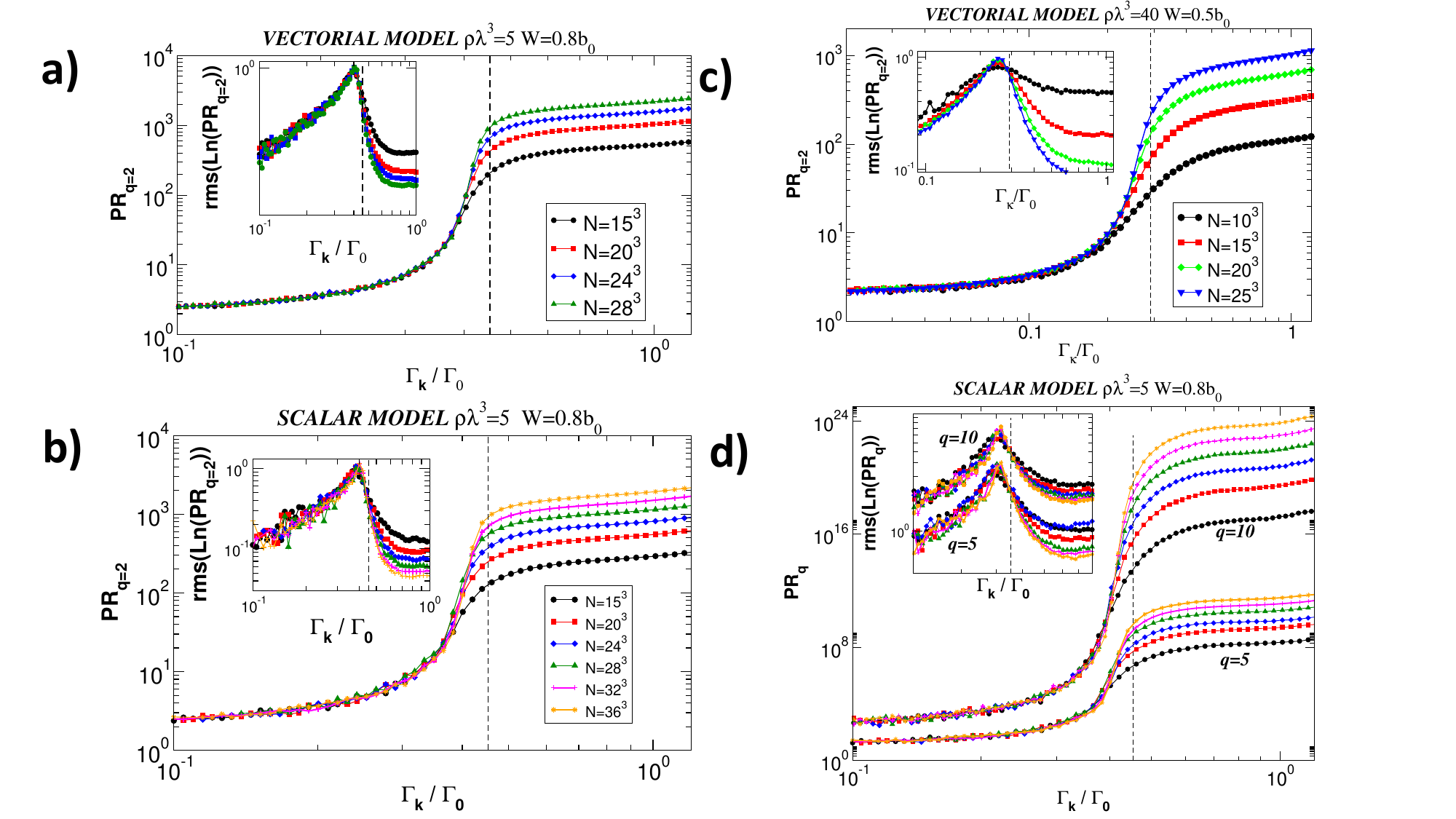} \\
\caption{ (Color online) {\it Mobility edge in the imaginary axis}. 
Typical GPR for $q=2$ (panel a,b,c) and $q=5,10$ (panel d) as a function of the decay width of the eigenstates are shown both for the vectorial (a,c) and scalar model (b,d) for different number $N$ of atoms and constant density, see legend. 
The vertical black dashed line indicates the critical width [Eq.~(\ref{Gcr})].  In the insets the root mean square of $\ln(PR_q)$ is shown  as a function of the decay width of the eigenstates.   In panels (a,b,d), for each $N$ the eigenvalues in the region
$-b_0/4<(E-E_0)/\Gamma_0<b_0/4$ were considered, while in panel c)the region
$-7-b_0/4<(E-E_0)/\Gamma_0<7+b_0/4$ were considered.
}
\label{fig:Fig2}
\end{figure*}

Note that $\mathcal{H}$ contains both real and imaginary parts, which
takes into account that the excitation is not conserved since it can leave the system by outgoing radiation. Its complex eigenvalues ${\cal E}=E-i \Gamma/2$ describe the energies and line-widths (decay rates) of the eigenmodes of
the system. We stress that even in the dilute limit $\rho\lambda^3\ll1$
we can have cooperative behaviour in the large sample limit ($L\gg \lambda$), provided that the cooperativity parameter is $b_0\gg 1$. In this regime cooperative effects such as single-excitation sub-
and superradiance become relevant~\cite{Kaiser12,Guerin16a,Guerin16b}.

In addition to the positional disorder of the atoms as studied previously \cite{Skipetrov14,Bellando14}, we now introduce an additional 
random diagonal disorder term in the Hamiltonian,  which shifts
the excitation energy of the atoms around its avarege value $E_0$. Such diagonal disorder terms have not been given sufficient consideration in the context of localization of light, as engineering such effects is difficult in typical condensed-matter samples. However in  cold atomic clouds, such on-site disorder can be realized by applying a speckle field coupling the excited state to an auxiliary excited state with convenient detuning, inducing thus random light shifts of the atomic resonances without inducing dipole forces in the ground state.
Following the approach of the Anderson model on a lattice, we allow the site energies to fluctuate
in the range of $[-W/2,+W/2]$, where $W$ is the strength of disorder. 
Ensemble averaging thus includes different realizations of the random position of the atoms and of site disorder. 
Within this model, we study both the eigenvalues as has been done in~\cite{Skipetrov14,Bellando14} as well as the eigenstates~\cite{Bachelard15}. 

\paragraph{Localized Subradiant States.} A  striking illustration of the existence of localized states  is given in Fig.~{\ref{fig:Fig1}}, where we represent a typical localized subradiant eigenstate for the scalar model.
The upper panel of Fig.~{\ref{fig:Fig1}} shows a 3D representation of a typical localized eigenstate, while the lower panel of Fig.~{\ref{fig:Fig1}} shows the projection of the squared wavefunction $|\psi(r)|^2$ on the $x-y$ plane.
While for zero diagonal disorder the vast majority of the states, which can be both superradiant or subradiant, are fully delocalized~\cite{sp} for the spatial density considered, adding sufficient diagonal disorder leads to localization of the longer-lived subradiant states.
We  observe that the localized peak, shown in the lower panel of Fig.~\ref{fig:Fig1}, comes hand in hand with an extended tail, thus exhibiting a hybrid character, in agreement with Refs.~\cite{Celardo16,Celardo13a,Celardo13b}.
We note that the presence of such extended tails might  strongly affect transport properties~\cite{chavez2021disorder}, for instance suppressing the exponential decay of transmission with the system size. Here we focus on the structure of the eigenmodes, leaving the analysis of the transport properties of subradiant localized states for a future work.

\begin{figure}[!htb]
\centering
\includegraphics[width=2.6in]{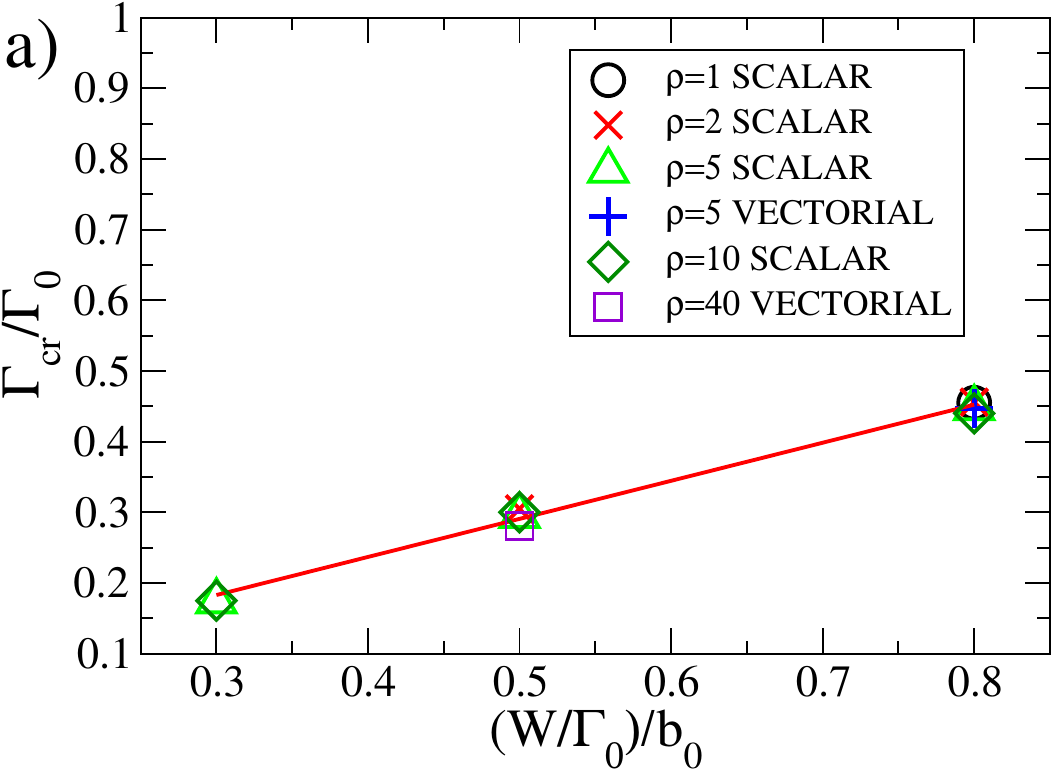} 
\includegraphics[width=2.9in]{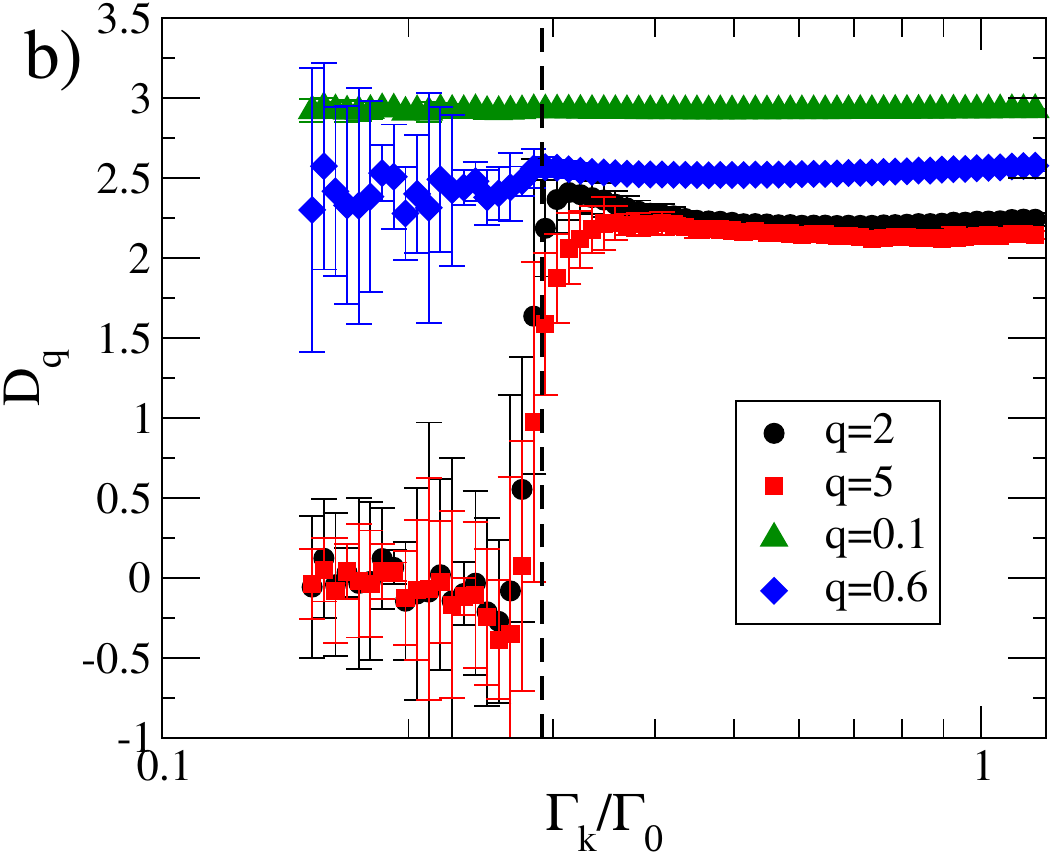} \\
\caption{(Color online)  {\it Panel a), critical decay width}: the critical decay width for the localization transition is shown as a function of the normalized disorder strength for different densities and for both the vectorial and the scalar model (see legend). The precision with which we determined the critical decay width is always below $\pm 0.015$. {\it  Panel b): fractal dimension}: the fractal dimension as a function of the normalized decay width is shown for the  case $\rho\lambda^3=5$,  $W/(b_0 \Gamma_0)=0.5$. The fractal dimension has been extracted from the size dependence of the GPR for different values of $q$. The vertical black dashed line indicates the critical width [Eq.~(\ref{Gcr})].}
\label{fig:Fig3}
\end{figure}

\paragraph{Mobility edge in the imaginary axis.}
In open systems, standard approaches to study localization such as the Thouless parameter should be applied with care~\cite{genack}. We therefore analyze the properties of the  generalized participation ratio (GPR) of the eigenfunction $\psi$ of the system~\cite{Rodriguez00,Rodriguez03}, 
\begin{equation}
 PR_q=  \left| \sum_i |\langle i| \psi \rangle|^{2} \right |^q /\sum_i |\langle i| \psi \rangle|^{2q}. 
\label{PRq}
\end{equation}

For localized eigenfunctions $PR_q$ is independent of the system size for all $q$, while in the delocalized regime  $PR_q \propto N^{q-1}$. On the other hand, at the localization transition the GPR diverges with $N$ as 
\begin{equation}
PR_q \propto N^{D_q(q-1)/d}
\label{PRq2}
\end{equation}
where $d$ is the embedding dimension and $D_q$ defines the fractal dimension. 
Moreover, the distribution $P(PR_q/PR_q^{typ})$, where $PR_q^{typ}=\exp{\langle \ln(PR_q)\rangle}$, is invariant at criticality in the large system size limit. This implies that the variance of the distribution $P(\ln(PR_q))$ is independent of the size at criticality~\cite{CuevasPRB2002,CuevasPRL2001,Mirlin2002,EversPRL2000,VargaPRB2002,BermudezEPL2012,BermudezJS2014,BermudezPRE2019}, allowing for a precise identification of the critical point. 

In order to have a  general view of the localization properties of the eigenmodes of our system, we computed the GPR of all the eigenmodes for a specific  disorder strength, and we plotted them as a function of their complex eigenvalues (real and imaginary part). A typical example of this analysis can be seen in   Fig.~\ref{fig:Fig2a}, which shows a strong dependence of the $PR_2$ of the eigenmodes on the imaginary part of their eigenvalues, while the dependence on the real part is weak.  Specifically we observe that the smaller is their imaginary part, the more the eigenmodes are localized. Note that the results of Fig.~\ref{fig:Fig2a} refer to the scalar model, and a similar figure for the vectorial model can be found in~\cite{sp}. These results are consistent with previous findings about the interplay of super- and subradiance with disorder~\cite{Celardo13a,Celardo13b,Giulio14,Giulio15}: subradiant states are the ones most affected by disorder.
The most interesting feature of this non-uniform response of the eigenmodes to disorder can be seen if one analyzes the typical value of $PR_q$ ($PR_q^{typ}$) as a function of the decay widths. Since the optical thickness $b_0$ sets a relevant energy scale of the system (\emph{i.e.}~the spectral energy broadening prior to adding the diagonal disorder)~\cite{Bellando14}, we considered different systems at a constant density and for a fixed value of the ratio $W/(b_0 \Gamma_0)$. The results are shown in Fig.~\ref{fig:Fig2}, both for the vectorial and the scalar model, for different system sizes at constant density. The results clearly indicate the presence of a transition in the behaviour of the GPR: 
while the typical $PR_{q}$ of the eigenmodes is independent of the system size below $\Gamma_{\rm cr}$ if $W/b_0$ is kept fixed (see vertical dashed line), it increases with the system size above $\Gamma_{\rm cr}$. These are precisely the same features present when analyzing the GPR of the 3D Anderson model (or other models displaying a localization transition) in correspondence of a mobility edge in the real energy. 
Thus our results points to the existence of a ``mobility edge'' in the imaginary axis. 
We checked that the imaginary mobility edge is independent of $E$ around the band center, as shown in Fig.~\ref{fig:Fig2a} and further discussed in~\cite{sp}.

We note that in the large density limit the results shown in Fig.~\ref{fig:Fig2}(c) are extremely interesting, since they indicate that in presence of diagonal disorder, a localization transition can exist even in the large density limit for the vectorial case in absence of any magnetic field. This is at variance with what has been stated in~\cite{Skipetrov14} where no diagonal disorder was considered. Moreover the mobility edge in the imaginary axis, even in the large density limit,  is well captured by Eq.~(\ref{Gcr}).

In order to identify the critical decay width corresponding to the imaginary mobility edge, we performed a systematic analysis of the variance of the GPR {\it vs} the disorder strength $W/\Gamma_0$ for
different densities, system sizes and ranges of decay widths. The variance of $\ln(PR_q)$ has been used in the literature to pinpoint the localization transition and it has been shown that, at the localization transition, the variance of $\ln(PR_q)$ is independent of the system size due to a universal distribution of the GPR~\cite{Mirlin2002}. Similarly to Ref.~\cite{BermudezPRE2019}, we use the crossing of rms$(\ln PR_q)$ close to its maximal value to locate the localization transition, see insets in Fig.~\ref{fig:Fig2}. This allowed us to identify a critical decay width $\Gamma_{cr}$. 

We studied the critical decay widths as a function of disorder for different densities. The results are shown in Fig.~\ref{fig:Fig3}(a). By fitting the numerical results we obtained 
 an expression for the critical decay width: 
\begin{equation}
\frac{\Gamma_{\rm cr}}{\Gamma_0} \approx 0.021+0.54 \frac{W}{b_0\Gamma_0}.
\label{Gcr}
\end{equation}
We note that the above expression cannot be extrapolated at small values of disorder since in that case the landscape of the GPR  can only be understood analysing the whole complex plane~\cite{sp}.

We have also analyzed the GPR for different $q$ values: $q=0.1,0.6,2,5,10$~\cite{sp}. For $q \ge 2$ we always find a clear signature of a localization transition  at a critical decay width, while for small values of $q$  a localization transition is not observed. This reflects the hybrid nature of the localized eigenstates: indeed together with a localized peak, an extended tail is present. The GPR for large values of $q$ is more sensitive to large  values of $|\psi|^2$, thus it describes the behaviour of the localized peak, whereas the GPR for small values of $q$ is sensitive to  small values of the wave function amplitudes and thus to the wave function tails. Since the tails are always extended (delocalized), no localization transition is seen for small $q$~\cite{sp}. 
In order to further confirm the above picture, we have computed the fractal dimension $D_q$ as a function of the decay widths.  The results are shown in Fig.~\ref{fig:Fig3}(b). As one can see for $q \ge 2$ a transition in the fractal dimension is seen from zero to a value larger than one, while for $q < 1$ no transition is observed, confirming the hybrid nature of the eigenmodes of the system. Note that in the extended phase, even for $q=2,5$, $D_q$ is different from $d=3$ indicating that the wave function are never fully extended.  In other words, the eigenfunctions are always multifractal both below and above criticality: $\Gamma_{\rm cr}$ marks the transition from a frozen phase (where the the $GPR$ is independent of $N$ for sufficiently large $q$), to a weakly multifractal phase (with a narrow distribution of fractal dimensions $D_q$)~\cite{Mirlin08}.

\paragraph{Conclusions.}We considered a well known radiative non-Hermitian Hamiltonian model  to describe coherent multiple scattering of light in cold atomic clouds at low excitation level. Our results give new insights on the problem of localization in open quantum systems under the interplay of non-Hermiticity and disorder. A novel kind of localization transition has been identified, occurring at a critical lifetime (or inverse decay rate $\Gamma$) of the eigenmodes of the system, i.e. along the imaginary energy axis. A single-parameter scaling was found for the critical decay rate $\Gamma_{cr}/\Gamma_0$ [Eq.~(\ref{Gcr})] for the localization transition, which is given by $W/(b_0 \Gamma_0)$, in contradiction to what could be expected from $\rho$ or $b_0$ separately. The localization transition identified here in a realistic model of light matter interaction shares many analogies with the Anderson transition in 3D lattices and with localization transitions in long-range interacting systems, such as in the power-banded random matrix model~\cite{Mirlin2002,Mirlin08}, but also important differences: the localization transition is signalled by the behaviour of the GPR for large $q$ values (larger or equal than 2) and not for small $q$ values (less than 2). We attribute this feature to the fact that the eigenmodes are not fully localized but that have a hybrid character, with a localized peak and an extended tail. A precise characterization of the shape of these eigenmodes will be the topic of a future work. Despite these differences, our results indicate the existence of a novel kind of localization transition occurring along the imaginary energy axis which is independent of the real energy (around the band center) for sufficiently large values of diagonal disorder and optical thickness. 
The existence of a mobility edge in the imaginary axis found in this Letter certainly constitutes a novel feature in the field of localization in open quantum systems.
Further research  will be necessary to assess the impact of our results. For instance the general conditions for this mobility edge in the imaginary axis to arise in open quantum systems should be investigated both in the single excitation and many excitation regime and for different topology and dimensions and critical exponents determined.  Our findings  
are relevant not only from a fundamental point of view but also for applications, \emph{e.g.}~to achieve efficient energy storage, quantum memory, quantum simulation and sensing devices.

\begin{acknowledgments}
We acknowledge stimulating discussions with R. Bachelard, A. Biella, F. Borgonovi, G. G. Giusteri and W. Guerin. 
Computing time was provided by the High Performance Computing Center of the University of Strasbourg. This research was also supported, in part, by the Center for Research Computing of the University of Notre Dame through access to key computational resources. Part of this work was performed in the framework of ERC Advanced Grant No. 832219 (ANDLICA). 
\end{acknowledgments}

\bibliographystyle{apsrev4-2}
\bibliography{references.bib}

\end{document}


\renewcommand{\theequation}{S\arabic{equation}} 
\renewcommand{\thefigure}{S\arabic{figure}} 

\title{Supplemental material for\\
``Localization of light in three dimensions:\\
a mobility edge in the imaginary axis in non-Hermitian Hamiltonians''}

\author{Giuseppe Luca Celardo}
\affiliation{Department of Physics and Astronomy, CSDC and INFN, Florence Section, University of Florence, Italy. European Laboratory for Non-Linear Spectroscopy, Università di Firenze, Via Nello Carrara 1, 50019 Sesto Fiorentino, Italy}
\author{Mattia Angeli}
\affiliation{John A. Paulson School of Engineering and Applied Sciences, Harvard University, Cambridge, Massachusetts 02138, USA}
\author{Francesco Mattiotti}
\affiliation{CESQ and ISIS (UMR 7006), University of Strasbourg and CNRS, 67000 Strasbourg, France}
\author{Robin Kaiser}
\affiliation{Universit\'e C\^ote d'Azur, CNRS, Institut
  de Physique de Nice, 06200 Nice, France}

\maketitle

\section{Extended subradiant state}
\label{supp1}

\begin{figure}[th!]
\centering
\includegraphics*[width=3.375in]{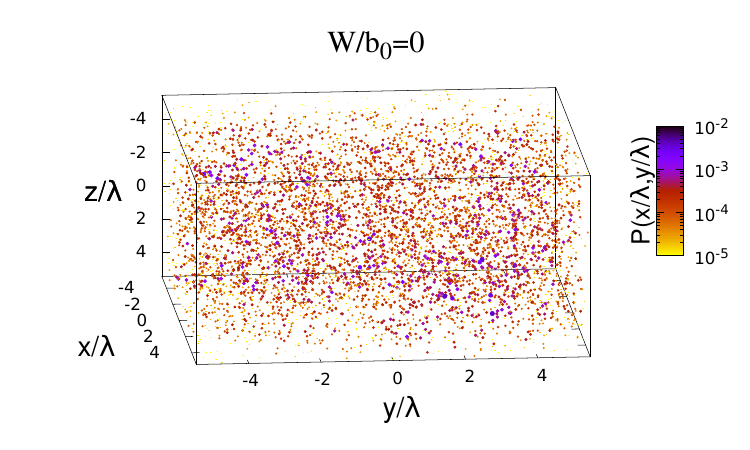} 
\includegraphics*[width=3.375in]{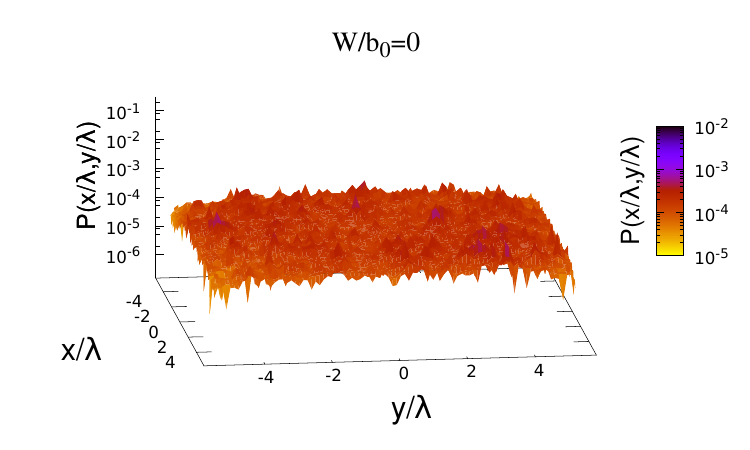} 
\caption{(Color online) Representations of a typical extended subradiant state (for the scalar model) similar to Fig.~1 in the main text. Upper panel: Three-dimensional representation of an eigenstate. The radius representing each atom is proportional to its excitation probability  $|\Psi_j(r)|^2$.
Lower panel: An eigenstate projected on the $x-y$ plane. Here $N=6400,\rho \lambda^3=5$ so that $b_0 \approx 17.3$ and $W/(b_0 \Gamma_0)=0$. 
For the state shown in both panels we have
$E/\Gamma_0=-0.0758, \Gamma/\Gamma_0=0.05$. 
The participation ratio $PR_2$, defined in Eq.~(5) of the main text, of the state shown in this figure is $PR_2=1941$.
}
\label{Fig1sup}
\end{figure}

Here we show an example of a typical \emph{extended} subradiant state for the scalar model in absence of diagonal 
disorder [$W/(b_0 \Gamma_0)=0$], see Fig.~\ref{Fig1sup}. This figure should be compared with Fig.~1 of the main text where a typical \emph{localized} subradiant state with $W/(b_0 \Gamma_0)=0.4$ is shown. Comparing the two figures one can see that   disorder 
in the transition frequencies of the atoms can induce localized states
in the subradiant subspace. Both in Fig.~\ref{Fig1sup} of this supplementary material and Fig.~1 of the main text, 
in the upper panels each atom is shown by a small sphere.  The probability  $|\Psi_j(r)|^2$ for the eigenstate to be on that atom is given by the color and the radius $R$ of the sphere according to the relation $R(r) = 1.5 (|\Psi_j(r)|^2/|\Psi_j(r)|^2_{max})^{2/7}$, where $|\Psi_j(r)|^2_{max}$ is the maximal probability for the case $W/(b_{0} \Gamma_0)=0.4$. This normalization relation was chosen to improve visibility. 
In the lower panels the projection on the $x-y$ plane of  $|\Psi_j(r)|^2$ on a grid of $60 \times 60$ is shown. To improve the quality of the representation,
each grid point has been averaged by the surrounding points, with a weighting inversely proportional to their distances squared.  

\section{Mobility Edge in the imaginary axis: scalar model}
\label{supp2b}

In order to analyze the localization transition we 
  consider the typical value of the Generalized Participation Ratio (GPR) of the eigenmodes of the system, see Eq.~(5) in the main text. We note that the eigenfunctions of the non-Hermitian Hamiltonian represent the projection of the total eigenfunctions on the single excitation manifold of the atomic degrees of freedom.   Thus the quantity $|\psi_k|^2$  which is used to compute the $PR_q$ represents the conditional probability to find the system on atom $k$, given that one quantum of excitation is stored in the system.  The state $|k\rangle$ is the state where the atom $k$ is excited while all the other atoms are in the ground state.

\begin{figure*}
\centering
\includegraphics[width=\textwidth]{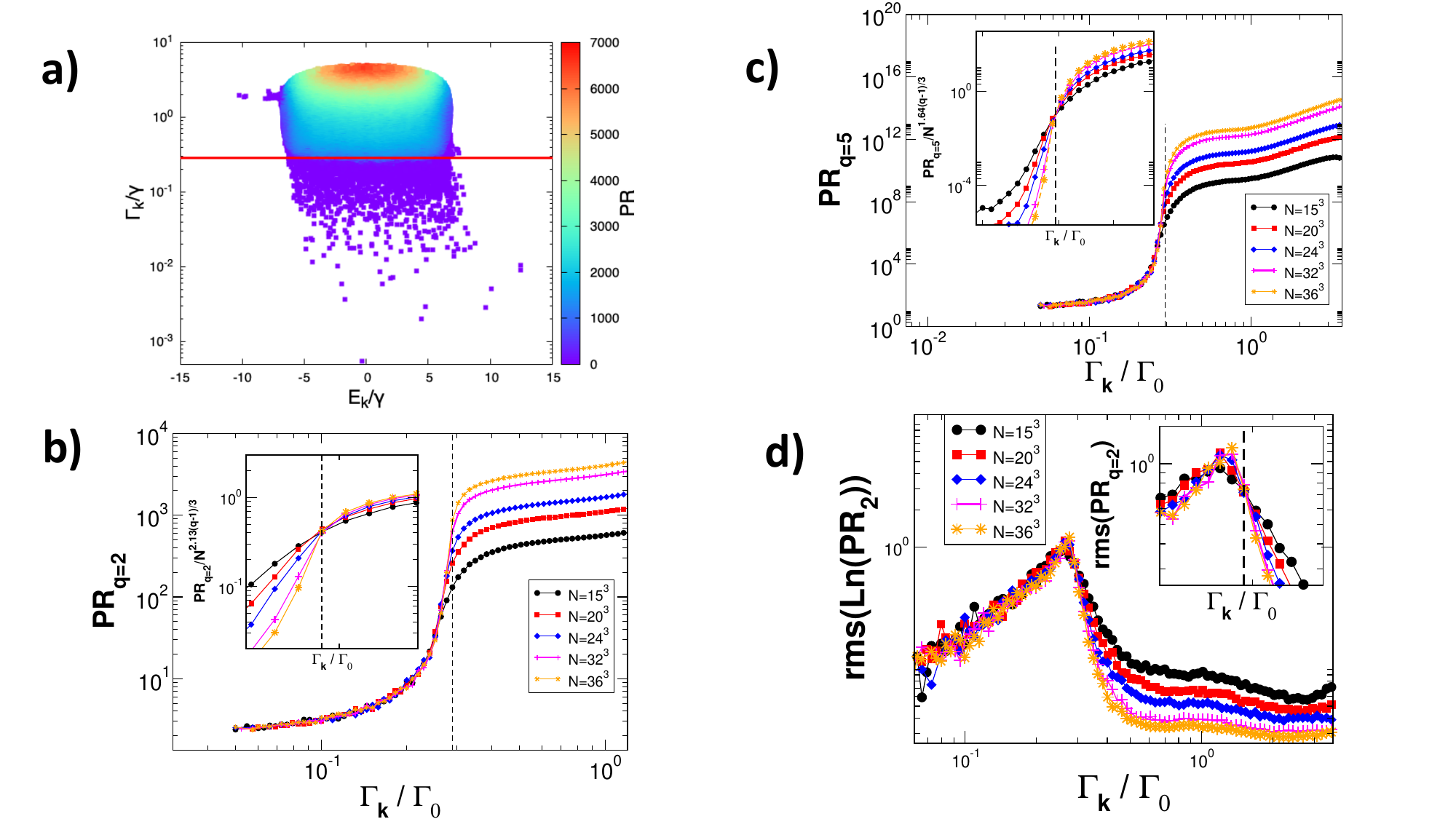} \\
\caption{(Color online) {\it Mobility edge in the imaginary axis}. 
(a) Participation ratio $PR=PR_{q=2}$ of the eigenstates in the complex plane $(E,\Gamma)$ of the eigenvalues of each state for $N=28^3$, $\rho\lambda^3=5$, $b_0 \approx 26.06$ and  $W/(b_0 \Gamma_0)=0.5$. Note that $E_k$ is the difference between the real part of the eigenvalues and $E_0$. 
The critical width for the transition to localization [Eq.~(7) in the main text] is indicated by the red  horizontal line. 
(b,c) Typical GPR for $q=2$ (panel b) and $q=5$ (panel c) as a function of the decay width of the eigenstates for $W/(b_0 \Gamma_0)=0.5$, $\rho\lambda^3=5$. 
The vertical black dashed line indicates the critical width [Eq.~(7) in the main text]. In the insets the $PR_q/N^{D_q(q-1)/d}$, where $D_q$ is the fractal dimension computed at criticality, is shown in the region around the localized-delocalized transition. {(d) Participation ratio fluctuations}. The root mean square of $\ln(PR_2)$ is shown  as a function of the decay width of the eigenstates. The inset shows an enlargement of the same panel around the transition. All panels refer to the case $W/(b_0 \Gamma_0)=0.5$, $\rho\lambda^3=5$ and different number $N$ of atoms, see legend. In panels (b,c,d), for each $N$ the eigenvalues in the region
$-b_0/8<(E-E_0)/\Gamma_0<b_0/8$ were considered.   
}
\label{fig:Fig2}
\end{figure*}

In order to have a  general view of the localization properties of the eigenmodes of our system, we computed the GPR of all the eigenmodes for a specific value of the disorder, and we plotted them as a function of their complex eigenvalues (real and imaginary part). A typical example of this analysis can be seen in   Fig.~\ref{fig:Fig2}(a), which shows a strong dependence of the $PR_2$ of the eigenmodes on the imaginary part of their eigenvalues, while the dependence on the real part is weak. Specifically we observe that the smaller is their imaginary part, the more the eigenmodes are localized. 
We also analyzed  the typical value of $PR_q$ ($PR_q^{typ}$) as a function of the decay widths for different systems at a constant density and for a fixed value of the ratio $W/(b_0 \Gamma_0)$.  The results are shown in Fig.~\ref{fig:Fig2}(b,c): they clearly indicate a localization transition.  
While the typical $PR_{q}$ of the eigenmodes is independent of the system size below $\Gamma_{\rm cr}$ if $W/(b_0 \Gamma_0)$ is kept fixed (see vertical dashed line), it increases with the system size above $\Gamma_{\rm cr}$. 
In order to identify the critical decay width corresponding to the imaginary mobility edge, we performed a systematic analysis of the variance of the GPR {\it vs.}~the disorder strength $W/\Gamma_0$ for
different densities, system sizes and ranges of decay widths. 
The variance of $\ln(PR_q)$ can be used, see main text, to pinpoint the localization transition: we use the crossing of rms$(\ln PR_q)$ close to its maximal value to locate the localization transition, see insets in Fig.~\ref{fig:Fig2}(d). This allowed us to identify a critical decay width $\Gamma_{\rm cr}$, see details in the main text. 
Using Eq.~(5) in the main text and performing a scaling analysis we can determine the fractal dimension as a function of the decay widths for different $q$ values: $D_q= \frac{d}{q-1}\ln (PR_q)/\ln(N)$. 
The rescaled typical GPR  $PR_q/N^{D_q(q-1)/3}$, where $D_q$ is the fractal dimension computed at criticality, is shown in the insets of Fig.~\ref{fig:Fig2}(b,c). As one can see the re-scaled $PR_q$ nicely cross at the critical decay width.

\begin{figure}
\centering
\includegraphics*[width=2.9in]{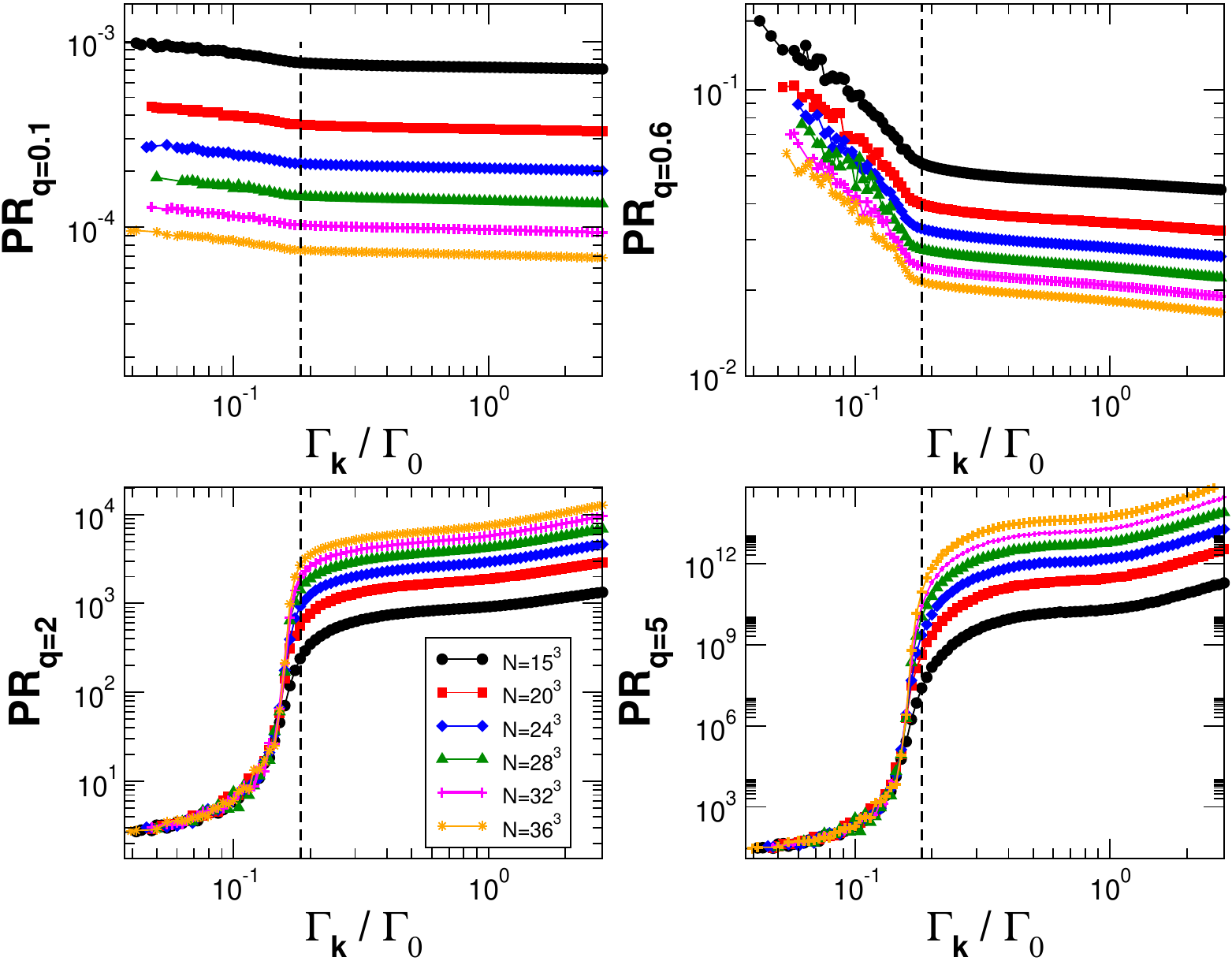} 
\caption{(Color online) {\it Mobility edge in the imaginary axis}. 
 Typical generalized participation ratio for $q=0.1,0.6,2,5$ as a function of the normalized decay width of the eigenstates for $W/(b_0 \Gamma_0)=0.3$, $\rho\lambda^3=5$. Here the typical $PR_q$ is averaged over the range
$-b_0/20<(E-E_0)/\Gamma_0<b_0/20+0.2$. 
The vertical black dashed line in all panels indicates the critical width obtained
from Eq.~(5) in the main text. Different numbers of atoms are considered: $N=10^3,15^3,20^3,24^3,28^3,32^3,36^3$.
}
\label{Fig3sup}
\end{figure}

In Fig.~\ref{Fig3sup} the typical value of $PR_q$ is shown for different values of $q$ for the case $\rho \lambda^3=5, W/(b_0 \Gamma_0)=0.3$. As one can see a clear signature of a localized-delocalized transition is shown for large values of $q=2,5$ (lower two panels), while for small values of $q=0.1,0.6$, $PR_q$ increases with the system size for all decays widths. As discussed in the main text, in the open quantum system considered here, localized eigenmodes have a hybrid nature,  with a localized peak and an extended tail. Small $q$ values are sensitive to the tails and thus reveal the extended character of the eigenmodes, while large $q$ values are more sensitive to the peak, thus revealing the localized character of the eigenmodes. Note that even if the $PR_q$ is never independent of the system size for small $q$ values, their dependence on the decay widths has a change of slope in correspondence of the imaginary mobility edge, see vertical dashed lines in the upper panels in Fig.~\ref{Fig3sup}. 

\begin{figure}
\centering
\includegraphics*[width=2.9in]{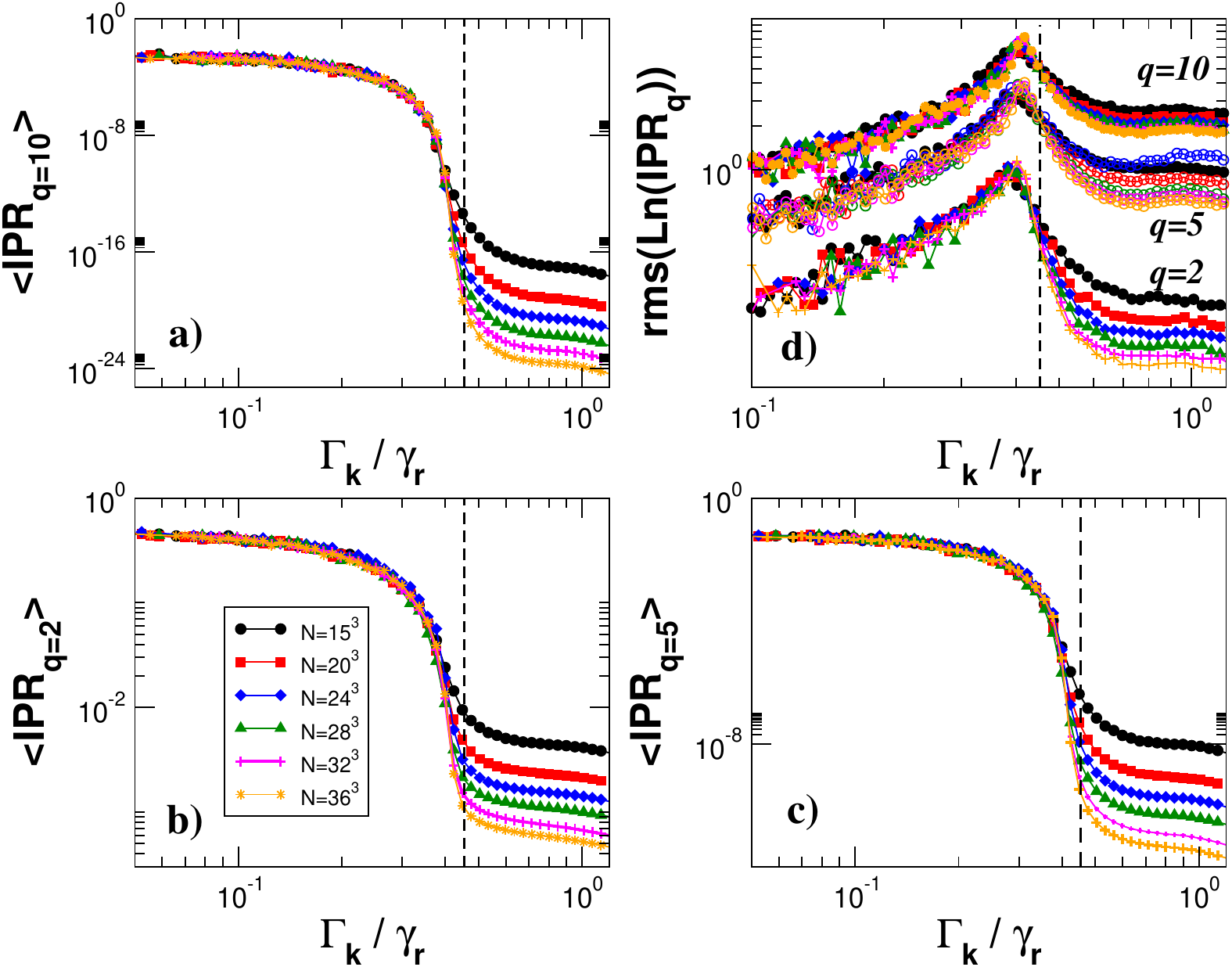} 
\caption{(Color online) {\it Mobility edge in the imaginary axis}. 
 Typical generalized inverse participation ratio for $q=2,5,10$ as a function of the normalized decay width of the eigenstates for $W/(b_0 \Gamma_0)=0.8$, $\rho\lambda^3=5$. Here the typical $IPR_q$ is averaged over the range
$-b_0/4<(E-E_0)/\Gamma_0<b_0/4$. 
The vertical black dashed line in all panels indicates the critical width obtained
from Eq.~(7) in the main text. Different numbers of atoms are considered: $N=10^3,15^3,20^3,24^3,28^3,32^3,36^3$.
}
\label{IPRsup}
\end{figure}

The mobility edge in the imaginary axis can be also analyzed considering the 
generalized inverse participation ratio (GIPR) of the eigenfunction $\psi$ of the system, 
\begin{equation}
 IPR_q=  \frac{ \sum_i |\langle i| \psi \rangle|^{2q} }{\left| \sum_i |\langle i| \psi \rangle|^{2} \right|^q }.
\label{PRq}
\end{equation}
For localized eigenfunctions $IPR_q$ is independent of the system size for all $q$, while in the delocalized regime  $IPR_q \propto N^{1-q}$. In Fig.~\ref{IPRsup}(a,b,c) the typical IPR is shown for $q=2,5,10$, showing  the mobility edge in the imaginary axis with the same critical decay width as computed in Eq.~(7) in the main text, see vertical dashed line. In Fig.~\ref{IPRsup}(d) the root mean square of $\ln(IPR_q)$ is shown for different $q$ values. As one can see our estimation for the critical decay width (see vertical dashed line) indicates fairly well the crossing of ${\rm rms}(\ln(IPR_q))$ for different $N$ even if a slight difference between the crossing for $q=2$ and $q=5,10$ is visible. A detailed investigation of this effect is beyond the scope of this manuscript and it will be investigated in future work.   Note that the root mean square of  $\ln(IPR_q)$ and  $\ln(PR_q)$ are the same since the inverse participation ratio and the participation ratio are just the inverse of each other.  

\begin{figure}
\centering
\includegraphics*[width=2.9in]{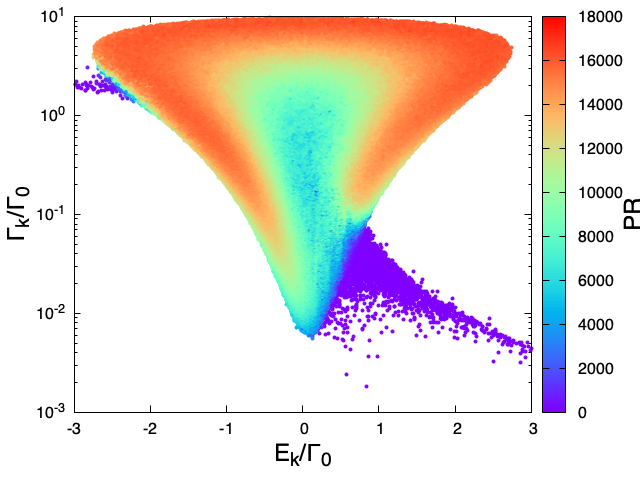} \\
\caption{(Color online) {\it Absence of a mobility edge in the imaginary axis in absence of disorder for the scalar case}. Participation ratio $PR=PR_{q=2}$ of the eigenstates (see legend on the right) in the complex plane $(E/\Gamma_0,\Gamma/\Gamma_0)$ of the eigenvalues of each state for $N=32^3$, $\rho\lambda^3\approx 5.05$, $b_0 \approx 30$ and $W=0$. Note that $E_k$ is the difference between the real part of the eigenvalues and $E_0$. 
}
\label{Fig2sup}
\end{figure}

Finally, it is important to note that our estimation of the critical decay width as a function of the disorder strength [Eq.~(7) in the main text] cannot be extrapolated to small values of disorder. Indeed for very small disorder the mobility edge in the imaginary axis is not defined, see Fig.~\ref{Fig2sup}. In general the  playground for the localization of open quantum systems is the complex plane, see Fig.~\ref{Fig2sup}, where the typical $PR_{q=2}$ for the case of zero disorder is shown in the complex plane. As one can see  comparing this figure with Fig.~2 of the main text and with Fig.~\ref{fig:Fig2}(a), for zero diagonal disorder no clear mobility edge in the imaginary axis is present. On the other side, one cannot exclude the presence of other mobility edges along different boundaries in the complex plane (this topic is outside the focus of the current manuscript and it will be investigated in a future work). 

\section{Mobility Edge in the imaginary axis: vectorial Model}

When considering the interaction of atoms with the electromagnetic field, 
the scalar model, see Eq.~(3) in the main text, is valid for dilute systems. In general the full vectorial character of light should be taken into account.

Here we describe in detail the radiative hamiltonian $\mathcal{H}_{\rm vec}$ describing light-matter interaction in atomic-like systems for weak fluence (single excitation approximation), see also~\cite{fra-nanoletter}.  
We  will focus on  atoms driven on a $s\rightarrow p$ dipole radiation transition which can be characterized by three degenerate levels in the excited state (labelled as $\alpha=x,y,z$), each with a transition dipole moment (TDM) equal in coupling strength and perpendicular to the others~\cite{robin-theory}.

Thus we model each atom as a four-level system, with a ground state $\ket{g}$ and three degenerate excited states $\ket{x}$, $\ket{y}$ and $\ket{z}$.
Corresponding TDM matrix elements are $\left\langle\alpha\left|\widehat{\vec{\mu}}\right|g\right\rangle=\mu{\hat{e}}_\alpha$, with $\alpha=x,y,z$ and the Cartesian unit vectors defined as ${\hat{e}}_\alpha$.

To model cold atomic clouds,
here we consider an ensemble of atoms randomly placed in a 3D box (positional disorder). Note that for the vectorial model the optical thickness has to be modified with respect to the scalar case, and we have: $b_0^{\rm (vec)}=(3/2) b_0^{\rm (scal)}$, where the optical thickness for the scalar case is given in Eq.~(4) of the main text. The Hamiltonian 
which takes the vectorial nature of light into account, describing an ensemble of atoms interacting with light can be written as~\cite{robin-theory}:
\begin{multline}
\label{hamvec}
  \mathcal{H}_{\rm vec} = \sum_{n=1}^N\sum_{\alpha\in\{x,y,z\}} \left( E_{n,\alpha} - i\frac{\Gamma_0}{2} \right) \ket{n,\alpha} \bra{n,\alpha} \\
  - \frac{\Gamma_0}{2} \sum_{\substack{m,n=1 \\ \left( m \neq n \right)}}^N \sum_{\alpha,\beta \in \{x,y,z\}} V_{m,n,\alpha,\beta}\ket{m,\alpha}\bra{n,\beta}~,
\end{multline}
where $E_{n,\alpha}$ are the atomic transition energies and $\Gamma_0$ is the radiative decay rate of a single atom. In Eq.~\eqref{hamvec}, $\ket{n,\alpha}$ represents a quantum state where the $n$th atom is excited in its $\alpha$th state, while all the other atoms are in their ground state.
Interaction terms are non-Hermitian, namely 
\begin{widetext}
\begin{align}
\label{hamc}
V_{m,n,\alpha,\beta} &= \frac{3}{2} e^{i k_0 r_{m,n}} \left[ \left( \frac{1}{k_0 r_{m,n}} + \frac{i}{k_0^2 r_{m,n}^2} - \frac{1}{k_0^3 r_{m,n}^3} \right) \hat{e}_\alpha \cdot \hat{e}_\beta - \left( \frac{1}{k_0 r_{m,n}} + \frac{3i}{k_0^2 r_{m,n}^2} - \frac{3}{k_0^3 r_{m,n}^3} \right) (\hat{e}_\alpha \cdot \hat{r}_{m,n}) (\hat{e}_\beta \cdot \hat{r}_{m,n}) \right]. 
\end{align}
\end{widetext}
In Eq.~\eqref{hamc},
$k_0=E_0/(\hbar c)$ is the transition wavenumber (with $E_0$ being the mean atomic transition energy $E_0=\braket{E_{n,\alpha}}$), $r_{m,n}$ is the distance between the $m$th and $n$th atom and ${\hat{r}}_{m,n}$ is the unit vector joining them. 
Note that this Hamiltonian has dimension $3N \times 3N$ contrary to the scalar case which had dimension $N \times N$. 

Using Eq.~(\ref{hamvec}) we have analyzed an ensemble of atoms in a 3D box occupying random positions. Diagonal disorder has been also considered allowing the energies  $E_{n,\alpha}$ 
 to fluctuate in the range
of $[-W/2, +W/2]$, where W is the strength of disorder. The typical values of the generalized participation ratio has been computed by computing the probability of the excitation to be on every atom. 

\begin{figure}
\centering
\includegraphics*[width=2.9in]{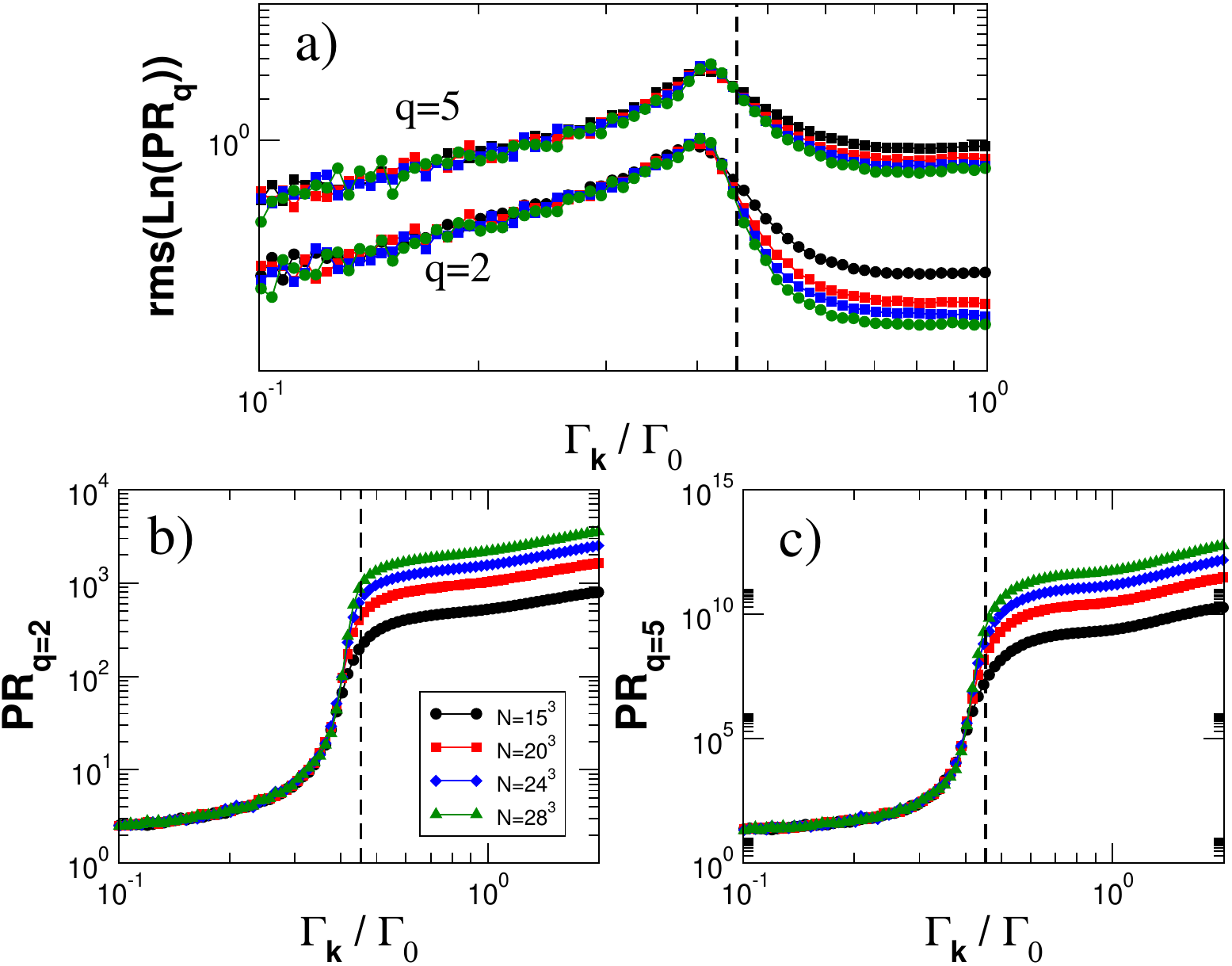} \\
\caption{(Color online) {\it Mobility edge in the imaginary axis for the vectorial case}. Panel (a): The root mean square of $\ln PR_q$ for $q=2,5$ is shown as a function of the decay width of the eigenstates for $W/(b_0\Gamma_0)=0.8$, $\rho\lambda^3=5$.
Panel b,c): Generalized Participation ratio for $q=2$ (panel b) and $q=5$ (panel c)  as a function of the decay width of the eigenstates for $W/(b_0 \Gamma_0)=0.8$, $\rho\lambda^3=5$. Here the typical $PR_q$ is averaged over the range
$-b_0/4<(E-E_0)/\Gamma_0<b_0/4$. 
The vertical black dashed line in all panels indicates the critical width obtained
from Eq.~(7) in the main text. Data in all panels refer to the vectorial model, see Eq.~(\ref{hamvec}).   
}
\label{Fig4sup}
\end{figure}

Fig.~\ref{Fig4sup} shows the results for the vectorial case $\rho \lambda^3=5, W/(b_0 \Gamma_0)=0.8$. As one can see, clear signatures of the mobility edge in the imaginary axis are shown  for the typical value of $PR_{q=2,5}$ and for rms$(\ln PR_{q=2,5})$. Note that our estimation of the critical decay width corresponding to the imaginary mobility edge given  in Eq.~(7) in the main text is in very good agreement with the numerical data, see vertical dashed lines in Fig.~\ref{Fig4sup}.

\begin{figure}
\centering
\includegraphics*[width=2.9in]{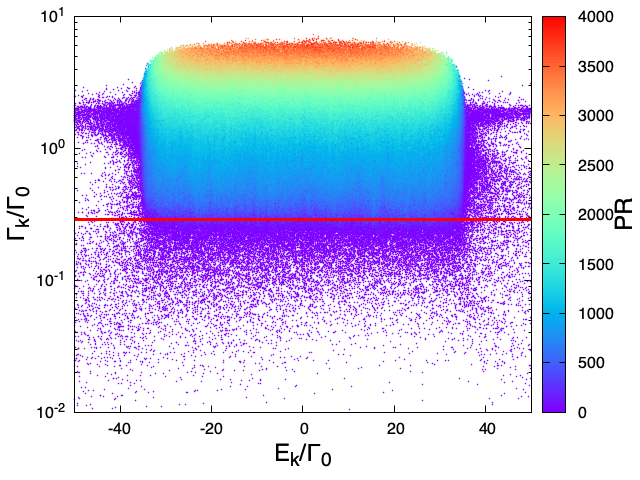} \\
\caption{ (Color online) {\it Mobility edge in the imaginary axis for the vectorial case}. Participation ratio $PR=PR_2$ of the eigenstates (see legend on the right) in the complex plane $(E/\Gamma_0,\Gamma/\Gamma_0)$ of the eigenvalues of each state for $N=25^3$, $\rho\lambda^3 = 40$, and $W/(b_0 \Gamma_0)=0.5$. The horizontal red line indicated the critical decay width, see Eq.~(7) in the main text.  Note that $E_k$ is the difference between the real part of the eigenvalues and $E_0$. }
\label{Fig5sup}
\end{figure}

\begin{figure}
\centering
\includegraphics*[width=2.9in]{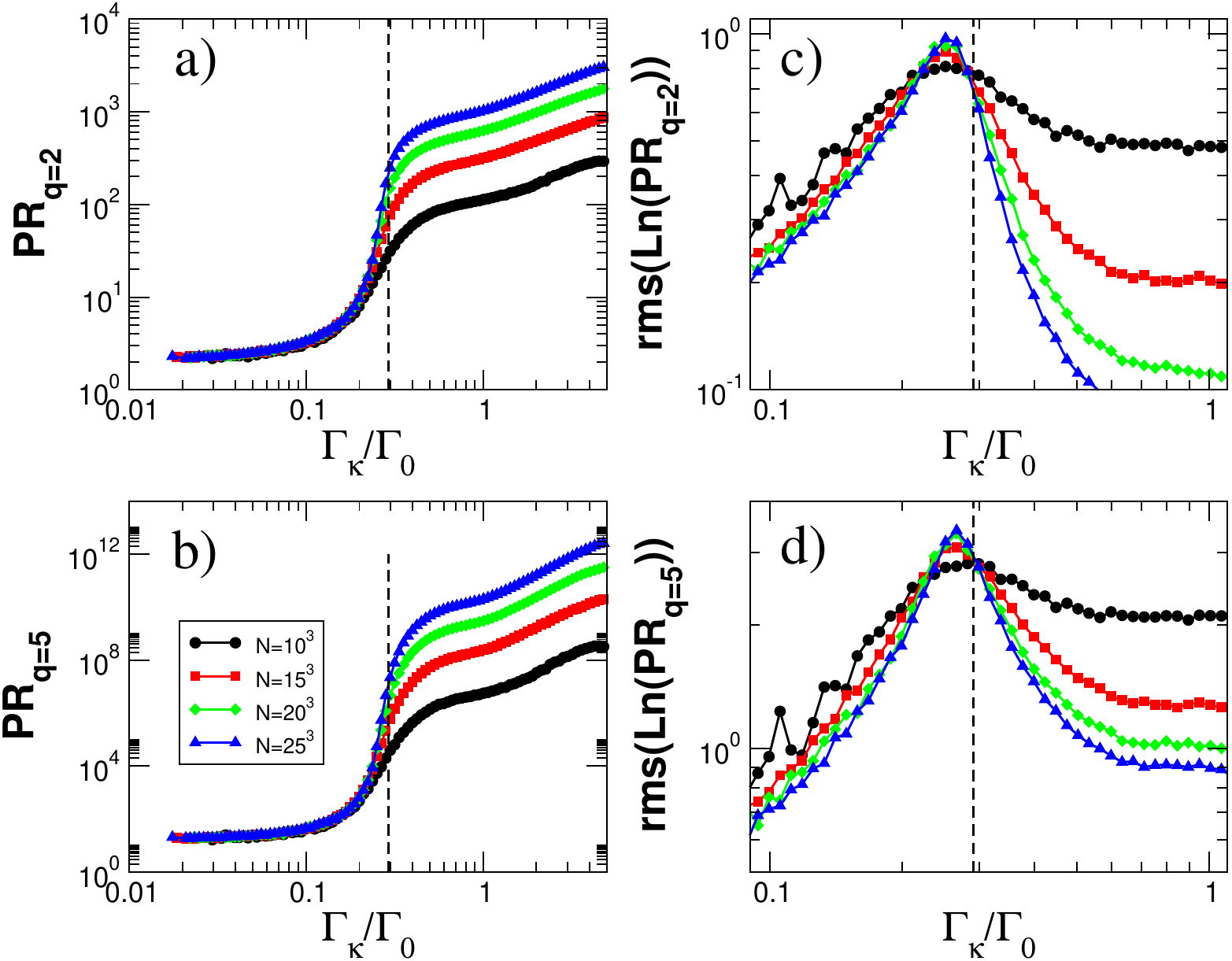} \\
\caption{(Color online) {\it Mobility edge in the imaginary axis for the vectorial case}. Panel (a,b): Typical generalized participation ratio for $q=2$ (panel a) and $q=5$ (panel b)  as a function of the decay width of the eigenstates. Panel (c,d): 
The root mean square of $\ln PR_q$ for $q=2,5$ is shown as a function of the decay width of the eigenstates.
Here the typical $PR_q$ is averaged over the range
$-7-b_0/4<(E-E_0)/\Gamma_0<7+b_0/4$. All data refer to the case  $W/(b_0 \Gamma_0)=0.5$, $\rho\lambda^3=40$. The vertical black dashed line in all panels indicates the critical width obtained
from Eq.~(7) in the main text. Data in all panels refer to the vectorial model, see Eq.~(\ref{hamvec}). 
}
\label{Fig6sup}
\end{figure}

Now we turn our attention to the large density case, where the full vectorial model is particularly relevant since the scalar model is a good approximation only in the dilute limit (small densities). It has been claimed that, in absence of a magnetic field, localization is not possible in the vectorial model of light~\cite{Skipetrov14}. 
Nevertheless, here we show that, even in the large density limit, the introduction of diagonal disorder induces a mobility edge in the imaginary axis which is well captured by Eq.~(7) in the main text, see Figs.~\ref{Fig5sup},\ref{Fig6sup}. 

\begin{figure}
\centering
\includegraphics*[width=2.9in]{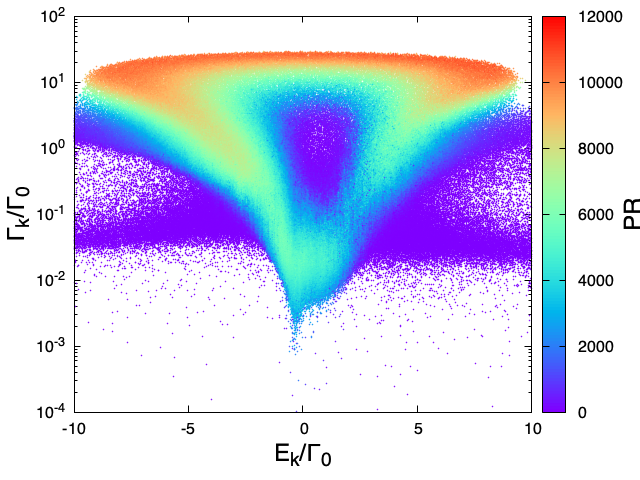} \\
\caption{(Color online) {\it Absence of a mobility edge in the imaginary axis in absence of disorder for the vectorial case}.  Participation ratio $PR=PR_{q=2}$ of the eigenstates (see legend on the right) in the complex plane $E/\Gamma_0,\Gamma/\Gamma_0$ of the eigenvalues of each state for $N=25^3$, $\rho\lambda^3 = 40$, and $W=0$. Note that $E_k$ is the difference between the real part of the eigenvalues and $E_0$.   
}
\label{Fig7sup}
\end{figure}

On the other hand, in absence of disorder and for large densities, the localized features of the eigenmodes of the system are more complicated to  capture. Again the real playground is the complex plane, see Fig.~\ref{Fig7sup}. The possibility to find localized states  even in absence of diagonal disorder, and in presence of positional disorder only, will be investigated elsewhere.

\section{On the nature of the mobility edge in the imaginary axis.}
Usually in open Anderson models~\cite{kottos}, the excitation can escape the system only from the boundaries,  so that 
the decay widths are proportional to the probability of a state to be
on the boundaries. As a consequence of this, most of the localized
states also have very long lifetimes (similar to subradiant states), 
since their probability to be on the
boundaries is exponentially small.  On the other side, the model
studied here, see also Ref.~\cite{biella},  strongly differs from the previously studied models of
localization in open systems, since in our case the excitation can
escape from any site and not only from the boundaries.
 For instance in our model a fully localized state on one site has a
 decay width equal to $\Gamma_0$, independent of the system size.

In order to clarify the difference between our model and previously studied open 3D Anderson models, let us consider a 3D cubic Anderson model with leads connected to one of its
 side as in Ref.~\cite{kottos}. Let us assume that the disorder is such to
 create a mobility edge at energy $E_c$. Clearly the decay width of
 the states will be very small for $E<E_c$, while they will be large
 for energy $E>E_c$, and correspondingly a mobility edge could also be
 found in the imaginary axis if one plots the participation ratio $PR_2$
 vs the decay widths. But in this case to use $E$ or $\Gamma$ is just
 a different way to label the states. On the other side our mobility
 edge has a completely different nature since it is independent of the
 real energy of the states in a wide energy range around the energy center, and it only depends on their imaginary energy, i.e. the lifetime of the eigenmodes of the open system. 
 
We note that the dependence of the $PR_2$ on the lifetime of the subradiant eigenmodes is a novel feature, which has not been captured by the toy model of Refs.~\cite{Celardo13a, Celardo13b}.
Indeed, in the  open 1D and 3D Anderson model  analyzed in
Refs.~\cite{Celardo13a, Celardo13b}, the sub- and superradiant modes
were segregated in two regions, whereas in the present case, no gap
between sub- and superradiant modes exists.

We also note that in the closed Anderson 3D model, the $PR_2$ diverges at a finite energy corresponding to the mobility edge.  In our case, the $PR_2$ diverges at a finite decay width (corresponding to the imaginary part of the complex eigenvalues of the system), thus we use the term mobility edge in the imaginary axis in analogy with the localization transition in closed systems which occurs along  the real axis. In the case of a closed system, such as the standard Anderson model, the behavior of the $PR_2$  reflects the transport properties of a system in a direct way: when the $PR_2$ increases with the system size, transmission will be diffusive or ballistic, while if the $PR_2$ is independent of the system size, transmission is exponentially suppressed with the system size due to localization. In the case of open systems, described by a  non-Hermitian Hamiltonian, the $PR_2$ has a more indirect link to transport properties since the eigenmodes are not fully localized but they have an hybrid nature as discussed in the main text. A discussion of the transport properties of hybrid states can be found in Ref.~\cite{disordernahum}. We do not aim to discuss this further in this manuscript, we just note that  $1/(E-\mathcal{H})$ is the propagator for the excitation in the system. For this reason, a change in the structure of the eigenmodes of $\mathcal{H}$ as signaled by the $PR_2$ represents a real physical change in the way excitations  propagate through the system.

\begin{figure}[ht!] 
\centering
\includegraphics*[width=3.375in]{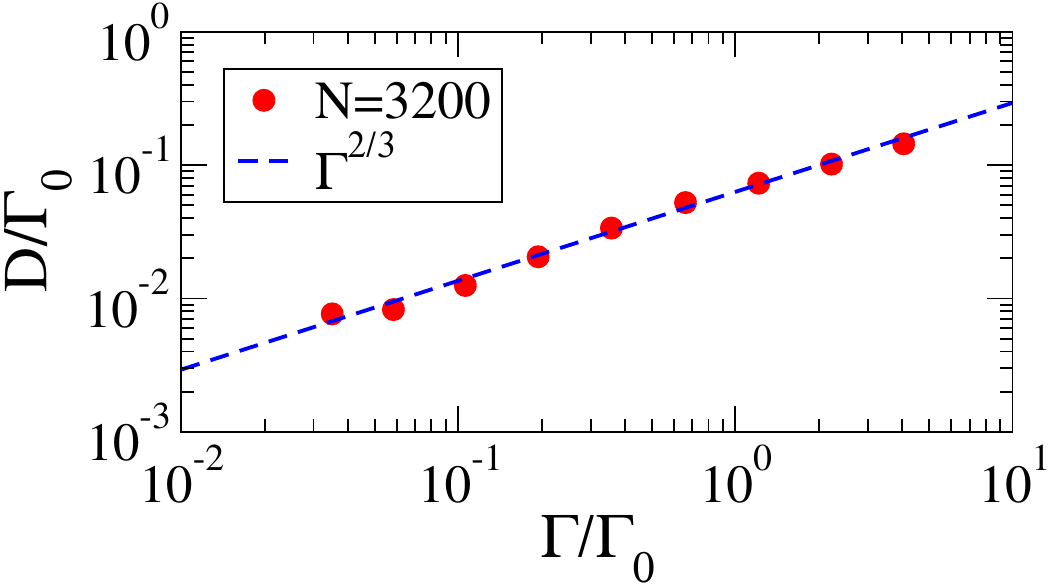} 
\caption{(Color online) {\it Increase of the  mean level spacing with the
    widths.} The mean level spacing in the complex plane $D$ is
  plotted {\it vs} the decay widths $\Gamma$ (red circle) for the case $N=3200$,
  $\rho\lambda^3=5$ and $W=0$. The mean level spacing has been computed by
  counting the number of complex eigenvalues per unit area in the
  complex plane for $-0.1<(E-E_0)/\Gamma_0<0.25$ and different ranges of $\Gamma$. 
  The mean level spacing $D$ has been obtained by taking the square
  root of the inverse density of complex eigenvalues. Note that an
  increase of the mean level spacing is observed even if one computes
  the distance in the real energy axis of the complex eigenvalues.
}
\label{DG}
\end{figure}

For the model considered here, the increase of the $PR_2$ with $\Gamma$  might be explained by the increase of the  mean level spacing of the eigenvalues with $\Gamma$, see Fig.~\ref{DG}. Note that in Fig.~\ref{DG} we compute the mean level spacing in the complex plane of the complex eigenvalues of the system. Indeed perturbation theory in the case of non-Hermitian Hamiltonian shows that it is the distance in the complex plane which determines the strength of perturbations~\cite{Giulio,nahum2}. Moreover we checked that also the mean level spacing in the real axis increases with the decay width. The increase of the mean level spacing with the decay width is due to the fact that  superradiant states have a stronger coupling to the photon field,  so that their energy spreads much more than subradiant states, which are partially shielded from the interaction~\cite{Celardo16}. Thus for a fixed amount of disorder, the states with lower $\Gamma$ are
more easily mixed by disorder than
the states with a larger $\Gamma$. 
Nevertheless this argument cannot explain the emergence of a mobility edge in the imaginary axis. More work is needed to deepen the understanding of this novel feature and to understand the general requirements for the mobility edge in the imaginary axis to emerge. 

We also note that a highy non-uniform  mean level spacing is typical for systems with long-range interactions. For instance  even a finite energy gap can be induced in such systems~\cite{Celardo16,nahum2}.
Localization even in presence of long-range interactions has been discussed for subradiant states in~\cite{Celardo13a,Celardo13b} and in general, in the framework of a shielding effect in~\cite{Celardo16,lea} and more recently in~\cite{deng}.

Finally we would like to point out that often in literature, localization properties are studied by means of the Thouless parameter~\cite{thouless77,Skipetrov14}.  
The Thouless parameter requires only the eigevalues of the system, nevertheless it  should be  used with care in open systems.
Indeed, in presence of absorption or other sources of leakage, 
the decay width of the states should be properly redefined in order to
take into account only the leakage from the boundaries, see discussion
in Ref.~\cite{genack}. Thus, analyzing the structure of the eigenmodes, as we did here,  is a much reliable mean to study localization in open quantum systems.

